\documentclass[sigconf]{aamas} 

\usepackage{balance} 

\setcopyright{ifaamas}
\acmConference[AAMAS '23]{Proc.\@ of the 22nd International Conference
on Autonomous Agents and Multiagent Systems (AAMAS 2023)}{May 29 -- June 2, 2023}
{London, United Kingdom}{A.~Ricci, W.~Yeoh, N.~Agmon, B.~An (eds.)}
\copyrightyear{2023}
\acmYear{2023}
\acmDOI{}
\acmPrice{}
\acmISBN{}

\newcommand{\truevalue}{\ensuremath{v^*}\xspace}
\newcommand{\linear}[1]{linear#1}

\renewcommand{\L}{\mathcal{L}}

\providecommand{\Mid}{\;\middle|\;}

\usepackage{multicol}
\usepackage{textgreek}
\usepackage[utf8]{inputenc}
\usepackage{amsmath}
\usepackage{natbib}
\usepackage{enumitem}
\usepackage{systeme}
\usepackage{tikz} 
\usepackage{xspace}

\usepackage{booktabs}
\usepackage{setspace}
\usepackage{arydshln}

\usepackage{cases}
\usepackage{makecell}
\usepackage[capitalize]{cleveref}
\usepackage{datetime}
\usepackage{algorithm}
\usepackage{algpseudocode}

\usepackage{booktabs, multirow} 
\usepackage{xcolor} 
\usepackage{changepage,threeparttable} 

\usepackage{subfiles}
\usepackage[capitalize]{cleveref}

\newcommand{\R}{\mathbb{R}}

\definecolor{darkred}{rgb}{0.7, 0.0, 0.0}
\definecolor{darkgreen}{rgb}{0.0, 0.7, 0.0}

\makeatletter
\newcommand*\bigcdot{\mathpalette\bigcdot@{.5}}
\newcommand*\bigcdot@[2]{\mathbin{\vcenter{\hbox{\scalebox{#2}{$\m@th#1\bullet$}}}}}
\makeatother

\newcommand*{\level}[1]{level-$#1$}
\newcommand{\quantalzero}{quantal-linear4\xspace}
\newcommand*{\qreplus}{QRE+L0\xspace}
\newcommand*{\qrep}[1]{QRE-$#1$\xspace}
\newcommand*{\nashapprox}{Nash approximation}
\newcommand*{\qchlinear}{QCH-QL4\xspace}

\newcommand*{\flzero}[1]{f^{L0}_{\text{#1}}}

\newcommand{\w}[1]{w_{\text{#1}}}
\newcommand{\f}[1]{f_i^{\text{#1}}}

\acmSubmissionID{118}

\title{Non-Strategic Econometrics (for Initial Play)}

\author{Daniel Chui}
\affiliation{
    Department of Computing Science
  \institution{University of Alberta / Amii}
  \city{Edmonton}
  \country{Canada}}
\email{dchui1@ualberta.ca   }

\author{Jason Hartline}
\affiliation{
Department of Computer Science
  \institution{Northwestern University}
  \city{Evanston}
  \country{USA}}
\email{hartline@northwestern.edu}

\author{James R. Wright}
\affiliation{
    Department of Computing Science
  \institution{University of Alberta / Amii}
  \city{Edmonton}
  \country{Canada}}
\email{james.wright@ualberta.ca  }

\begin{abstract}
  Modelling agent preferences has applications in a range of fields including economics and increasingly, artificial intelligence. These preferences are not always known and thus may need to be estimated from observed behavior, in which case a model is required to map agent preferences to behavior, also known as structural estimation.  Traditional models are based on the assumption that agents are perfectly rational: that is, they perfectly optimize and behave in accordance with their own interests. Work in the field of behavioral game theory has shown, however, that human agents often make decisions that are imperfectly rational, and the field has developed models that relax the perfect rationality assumption. We apply models developed for predicting behavior towards estimating preferences and show that they outperform both traditional and commonly used benchmark models on data collected from human subjects. In fact, Nash equilibrium and its relaxation, quantal response equilibrium (QRE), can induce an inaccurate estimate of agent preferences when compared against ground truth.  
  
  A key finding is that modelling non-strategic behavior, conventionally considered uniform noise, is important for estimating preferences.  To this end, we introduce quantal-linear4, a rich non-strategic model.  We also propose an augmentation to the popular quantal response equilibrium with a non-strategic component. We call this augmented model QRE+L0  and find an improvement in estimating values over the standard QRE. 
  QRE+L0 allows for alternative models of non-strategic behavior in addition to quantal-linear4.
\end{abstract}


\newcommand{\BibTeX}{\rm B\kern-.05em{\sc i\kern-.025em b}\kern-.08em\TeX}

\begin{document}


\pagestyle{fancy}
\fancyhead{}


\maketitle 


\section{Introduction}

This paper contributes to the study of behavioral models for initial
play and of structural inference of preferences from
behavior in games.
The former has shown that rich, parameterized models of behavior, including non-strategic behavior, give better predictions than more classical equilibrium models.
The latter uses strong equilibrium assumptions to
estimate preferences from equilibrium behavior.
We conduct experiments that show preferences and behavioral
models can be simultaneously inferred from initial play.


A main application of structural inference is in counterfactual
estimation.  After inferring preferences from behavioral data,
counterfactual scenarios can be evaluated.  Such evaluation can be
used, for example, in mechanism design for optimizing over many
mechanisms to find the one with the best equilibrium performance. A challenge for structural inference is that its
predictions are only guaranteed to be accurate if the assumed model is
correct.  In contrast, randomized controlled trials -- called A/B
testing by technology firms -- can directly evaluate a novel
mechanism, but to optimize over many mechanisms the sample size that
can be allocated to each mechanism is small.  \citet{CHN-16}, for
example, showed that methods from structural inference have an
exponential improvement for sample complexity in mechanism design over
randomized controlled trials.  However, this improvement comes with the aforementioned
reliance on the accuracy of the model.


The literature in behavioral game theory considers models of behavior
that relax the strong notion of equilibrium of classical game theory.
Especially for initial play, i.e., for behavior of players who do not
have prior experience playing a given game, classical notions of
equilibrium are bad predictors of behavior while behavioral models,
such as those in the quantal cognitive hierarchy (QCH) family, are good
predictors \cite{camerer2004cognitive,mckelvey1995quantal}.  The QCH
model combines quantal response (i.e., choosing actions with
probability proportional to the exponentiated payoff of the action)
with cognitive hierarchy (i.e., with levels of strategic thinking
and agents at each level responding only to those at lower levels and
with \level{0} corresponding to non-strategic behavior).
Recently, \citet{wright2019level} showed that rich \level{0}
models significantly improve accuracy of predicted behavior in QCH.


This paper develops behavioral models and conducts experiments within the context of 
initial play to measure the accuracy of these models in predicting
behavior and inferring preferences.  The experiments again highlight
the importance of rich \level{0} models in modeling behavior in
initial play.  The classical model of \level{0} behavior is uniform
randomization.  Uniform randomization ignores payoffs and is thus
unhelpful for inferring preferences.  Our results on predicted
behavior reinforce those of \citet{wright2019level}, showing that rich
models of \level{0} behavior are better predictors.  Moreover, and
intuitively, these models take into account payoffs and, thus, the
inferred \level{0} behavior aids in the inference of preferences. 
In fact, we find that the \level{0} model drives most of the gains in predictive behavior and inferring preferences, and the choice of strategic model is not as important.


Our experimental analysis introduces a new \level{0} model which is
derived from adding quantal response to the \linear4 \level{0} model
from \citet{wright2019level}.  The model that best predicts behavior
and admits the most accurate inference of values is quantal
cognitive hierarchy with this quantal \linear4 \level{0} model.  We
compared this model with the classical equilibrium and behavioral
models without rich \level{0} behavior of Nash equilibrium, quantal
response equilibrium \cite{mckelvey1995quantal}, and quantal cognitive
hierarchy (with uniform \level{0} behavior).  We also considered
quantal response equilibrium augmented with the a non-strategic model, including the aforementioned \quantalzero model.  Our models outperform these classical models with Nash equilibrium being the worst at both inference and prediction.

Our experimental setup considered 3-by-3 bimatrix games with randomly
generated payoffs.  This family of games is commonly studied in the
behavioral game theory literature \cite[e.g.,][]{mckelvey1995quantal, noti2021behavioral}.
We assumed payoffs were derived
from the classical single-dimensional linear model of auction theory
where payoffs are given linearly as a value for units of a good (i.e.,
an allocation) and a payment \cite{mcfadden1981econometric} (Our games allow payments to be
negative, i.e., some payoffs are given by some units of the good and a
negative amount of money.).  A key simplification of our game design is
that the players in our experiments were only aware of the payoffs in
the game and not of the decomposition of those payoffs into allocation
and payments.  Thus, we do not see in our data behavioral artifacts
related to whether or not the players can do the utility calculations
from allocations and payments.  Moreover, with such a design we are
free in our analysis to consider counterfactual inference questions
with various decompositions of payoffs into allocations and payments.

\section{Related Work}
The task of inferring preferences from observed data has generally been studied under the game theoretic assumption that players are in equilibrium  \citep[e.g.][]{athey2010structural,guerre2000optimal,athey2007nonparametric,paarsch2006introduction}.
In cases where the equilibrium assumption has been relaxed, this has generally been under the condition of repeated play (i.e., the subjects play the same game(s) repeatedly).
\citet{crawford2007level} and \citet{goeree2002quantal} use non-equilibrium behavioral models (level-$k$ thinking) to explain a widely-observed behavioral phenomenon---overbidding in private-value auctions---that is inconsistent with the bidders' being in equilibrium.  However, their experimental evaluation focuses on estimating parameters of the behavioral model only, taking the values as known to the analyst.
\citet{nekipelov2015econometrics} estimate private values from auction data without equilibrium assumptions, instead relying on a weaker assumption that agents use some form of no-regret learning.
 Similarly, \citet{ling2018game} provide a framework to learn game parameters from actions in zero-sum games, but do not validate their results on empirical data.

The work that most closely resembles our own is that of \citet{noti2021behavioral}, which has a similar objective of using models from behavioral game theory to infer preferences from empirical data in normal form games where the values are known but hidden from the analyst.
Our work differs in one key aspect, however; whereas Noti attempts to estimate values using player responses over repetitions in a single game, agents in our scenario only see each game once.
None of these aforementioned works study value estimation in initial play; each either relies upon repetition across games, and/or does not estimate values at all.





Value estimation has also been studied empirically under conditions resembling initial play in the field of school matching.  Value estimation is necessary for counterfactual evaluation of mechanisms and there are several papers \cite[e.g.][]{calsamiglia2020structural, hwang2015robust, he2015gaming, agarwal2018demand} which attempt to infer preferences of agents to evaluate the welfare of alternative mechanisms.  The way in which preferences are modelled vary between an equilibrium model to assuming all agents use simple behavioral rules.  
Notably, \citet{calsamiglia2020structural} construct a model of strategic and non-strategic agents in which strategic agents best-respond noisily to all other agents, including non-strategic agents, similar in principle to \qreplus.
Whereas non-strategic agents directly report their true preferences in the school choice setting, our framework allows us to consider scenarios where an indirect mechanism maps the preferences of non-strategic agents to actions.

Behavioral game theory aims to predict empirical human behavior better than traditional game theoretic concepts such as Nash equilibrium.
One well known model, quantal response equilibrium \citep{mckelvey1995quantal}, relaxes the strict optimization assumption made by Nash equilibrium, while maintaining the assumption that agents mutually respond to each others' strategies.
In contrast, iterative behavioral models such as level-$k$ models \citep{stahl1994experimental} and cognitive hierarchy \citep{camerer2004cognitive} assume that agents perform a fixed number of iterations of strategic reasoning, starting from a default strategy called the \level{0} strategy.  \citet{wright2017predicting} found that a combination of the two approaches, quantal cognitive hierarchy, performs best at predicting actual initial play in human subject experiments.  In later work, they showed that prediction performance can be further improved by specifying parameterized \level{0} models that combine simple decision rules, instead of the uniform randomization specification that is most frequently studied \cite{wright2019level}.

Another work that focuses on human behavior in initial play for normal form games is that of \citet{fudenberg2019predicting}.  Starting from the premise that initial play is reasonably approximated by \level{1} of iterated reasoning, they algorithmically generate games that are not captured well by \level{1} reasoning and construct a decision tree based model that improves on the prediction of the modal action in normal form games over that of previous economic models. A key point to note is that their \level{1} response relies on uniform randomization by non-strategic (\level{0}) agents. Further work by the same authors \citep{fudenberg2019measuring, fudenberg2020flexible} continues to rely on this uniform assumption. 
Our work focuses on inferring preferences of the agents and so exploits a richer model of \level{0} behavior in which non-strategic agents are sensitive to their own preferences. 
\section{Preliminaries and Setup}
The analyst's objective is to predict how much participants value a unit of some good, given their behavior in a set of normal form games.
Each player $i$ receives both an allocation of $x_i$ units of the good, and a payment $p_i$ in currency.
We assume that player $i$'s utility is linear in both payments and allocations;
i.e., player $i$'s utility is
$u_i(x,p) = vx_i + p_i$, where $v \in \R$ is $i$'s value (in currency units) for each unit of the good.
In this work, we will assume that the valuation $v$ is common across all players.

It is challenging to translate this setting into an experiment.
The main challenge is that we need to endow our experimental participants with a specific value for the good, which is common knowledge across all participants.
Presenting participants in the experiment with a valuation is not sufficient:  participants may not believe the valuation presented to them (i.e. they will believe the purpose of the study is something other than what is presented to them), or behavioral issues (e.g., arithmetic errors) may arise. 

To resolve these issues, we translate our setting into an experiment in a slightly less direct way.
Instead of presenting the outcomes of a game as a decomposition of units of good allocated and units of payment to the participant, we instead present the induced utilities.  That is, we perform the arithmetic for the participants of converting an allocation and payment to a utility.
We then map the behavior observed in these translated games (which we refer to as \emph{payoff games}) to a utility-equivalent decomposed game (which we refer to as \emph{allocation games}) and perform our analysis as if the players had chosen their actions in the allocation games.
Notice that, for any given payoff game, any number of utility-equivalent allocation games can be constructed.
As we will see, this allows us to repeat our analysis for different games and even different valuations using the same dataset of observations.  
We provide a more detailed explanation in Section \ref{sec:game_construction}.

\begin{table}[h]\centering
\caption{A summary of the strategic and non-strategic components included in our evaluation and the parameters $\theta$ for each component.}\label{tab:models_compared }
\begin{tabular}{lrrrr}\toprule
Strategic Component &$\hat{\theta}_S$ &Non-strategic Component &$\hat{\theta}_{NS}$ \\\midrule
Nash & $\varnothing$ &none & $\varnothing$ \\
QRE &$\beta, \lambda$ &uniform randomization & $\varnothing$ \\
PQCH &$\tau, \lambda$ &\quantalzero&$w_{L0}, \lambda_{0}, $ \\
None & $\varnothing$ & & \\
\bottomrule
\end{tabular}
\end{table}

\subsection{Behavioral Models}\label{section:behavioral_models}



Our behavioral models combine a component modelling strategic behavior with a component that models non-strategic behavior. 
At a high level, strategic behavior is that which responds to the anticipated actions of other agents while non-strategic behavior does not. 
The strategic models we consider are: Nash equilibrium, quantal response equilibrium (QRE), quantal cognitive hierarchy (QCH), and no strategic behavior.  
Of the strategic models, Nash and QRE are equilibrium models while QCH is not.
The non-strategic models we consider are: uniform randomization, quantal
\linear4 (QL4), and no non-strategic behavior.  These non-strategic models satisfy the formal definition of non-strategic behavior given in \citet{wright2022formal}.
With the exception of Nash equilibrium, each of the components just described have free parameters that must be learned from the data.  We refer to these parameters as ``behavioral parameters'', to distinguish them from the valuation parameters describing agent preferences that we also estimate from data.
Table \ref{tab:models_compared } summarizes the behavioral parameters for each component included in our main evaluation.

\subsection{Strategic Models}

The strategic models of Nash equilibrium, quantal response
equilibrium, and quantal cognitive hierarchy are defined
formally below.

\begin{definition}[Nash equilibrium]
    Let $BR_i(s_{-i})=\{s_i \in \Delta(A_i) \mid u_i(s_i,s_{-i}) \ge u_i(s'_i,s_{-i}) \forall s_i'\in\Delta(A_i)\}$
    be the set of best responses to $s_{-i}$.  Then a mixed strategy profile $s$ is a \emph{Nash equilibrium} if every agent $i$'s mixed strategy $s_i$ is a best response to the profile $s_{-i}$ of mixed strategies of the other agents: $s_i \in BR_i(s_{-i})$.
\end{definition}

The non-Nash equilibrium models that we consider are based on a relaxation of best response called quantal best response (QBR), in which agents play higher-utility strategies with higher probability (rather than strictly maximizing).  QBR is parameterized by a \emph{precision} (denoted by $\lambda$), indicating agents' sensitivity to utility differences.

\begin{definition}[Quantal best response]
  Let $u_i(a_i,s_{-i})$ be agent $i$'s expected utility when playing action
  $a_i \in A_i$ against mixed strategy profile $s_{-i}$ in game $G$.  Then a \emph{quantal
  best response} $QBR_i(s_{-i};G,\lambda)$ by agent $i$ to $s_{-i}$ is a mixed
  strategy $s_i$ such that
  \begin{equation}
  s_i(a_i) =  \frac{\exp[\lambda\cdot u_i(a_i, s_{-i})]}{\sum_{a'_i \in A_i}\exp[\lambda\cdot u_i(a'_i, s_{-i})]}.\label{eq:qbr}
  \end{equation}
\end{definition}

\begin{definition}[Quantal response equilibrium]
    A strategy profile $s$ of a game $G$ is a \emph{quantal response equilibrium} (QRE) with precision $\lambda>0$ when each agent quantally best responds to the strategies of the other agents;
    that is, when $s_i = QBR_i(s_{-i};G,\lambda)$ for all agents $i\in N$.
\end{definition}

Quantal cognitive hierarchy (QCH) is a non-equilibrium model, in which agents are heterogeneous in the number of steps of strategic reasoning they can perform.  Higher-level agents choose their actions in response to the strategies of lower-level agents.  The lowest level agents (level-0 agents) choose their actions non-strategically; that is, without reasoning about the actions of the other agents.
Level-0 agents are commonly specified to simply play a uniform distribution over actions; we evaluate that specification, but we also evaluate QCH using a richer specification of level-0 behavior (see Section~\ref{section:non_uniforml0}, below).

\begin{definition}[Quantal cognitive hierarchy]
    Quantal cognitive hierarchy with precision $\lambda>0$, level distribution $L$, and level-0 specification $f$ specifies that each agent $i$ has a level $k_i \sim L$.
    Let $\pi_{i,k} \in \Delta(A_i)$ be the distribution over actions predicted for an agent $i$ with level $k$.
    Level-0 agents play actions according to $\pi_{i,0} = f(G)$, where $f$ is some non-strategic function of the game payoffs.
    Agents with level $k>0$ play according to the distribution $\pi_{i,k} = QBR_i(\pi_{-i,0:k-1};G,\lambda)$, where
    \[\pi_{i,0:k}=\frac{\sum_{\ell=0}^kL(\ell)\pi_{i,\ell}}{\sum_{\ell'=0}^kL(\ell)}\]
    is the distribution over actions induced by conditioning on the level being at most $k$.

    The overall distribution of actions predicted by quantal cognitive hierarchy is
    $\pi_i = \sum_{k=0}^\infty L(k)\pi_{i,k}$.
\end{definition}

For the distribution of levels in QCH, a Poisson distribution is commonly used \cite[e.g.][]{camerer2004cognitive,fudenberg2020flexible}. We do the same and estimate a mean parameter $\tau$ on a truncated Poisson distribution where the max level of an agent is 3.
\begin{definition}[Poisson quantal cognitive hierarchy] 
Poisson quantal cognitive hierarchy is a specification of QCH in which the level distribution $L$ is specified by a Poisson distribution with the mean parameter $\tau$:
\begin{align*}
    L_{\tau; 0:k}=\sum_{\ell=0}^{k=3}\frac{\text{Poisson}(\ell;\tau)}{\sum_{\ell'=0}^k\text{Poisson}(\ell';\tau)}
\end{align*}

where $L_{\tau;\ell}$ is the proportion of agents at level $\ell$ given mean $\tau$ and with  $L_{\tau;0:k}$ sums to 1.
\end{definition}

\subsection{Non-strategic Models}\label{section:non_uniforml0}

It is standard in the literature to assume that non-strategic agents
randomize uniformly over their actions.  Recently,
\citet{wright2019level} found that using a linear combination of simple
decision rules as a level-0 specification markedly improves the prediction performance of QCH.
In this model, called \linear4, each decision rule identifies an action from $A_i$ that optimizes some simple criterion (e.g., maximizing the sum of all players' utilities), and predicts that player $i$ will play that action.\footnote{In the case of ties, the decision rule predicts a uniform distribution over the criterion-optimizing actions.}
The predictions of the simple decision rules are then linearly combined into an overall prediction, using weights that are free parameters of the model.

We evaluate a level-0 model adapted from \linear4 that we refer to as \quantalzero.
The key difference between the two models is that in \quantalzero, each decision rule computes its prediction as a quantal response to the different actions' criterion values.
In contrast, the predictions for  \linear4 are computed using strict optimization---each decision rule assigns probability 0 to each action that does not optimize its criterion.
This extension is motivated by two considerations.
%
First, behavioral models that assume quantal response to preferences have tended to predict better than equivalent models based on strict optimization: QRE predicts better than Nash equilibrium, QCH predicts better than cognitive hierarchy, and the level-$k$ model using quantal response predicts better than level-$k$ using best response \cite{wright2017predicting}.  It is thus natural to expect that modeling non-strategic agents as responding quantally will also improve prediction performance.
%
Second, the likelihood for \linear4 is continuous in the weights of the decision rules (i.e., in its behavioral parameters), but discontinuous in the valuation parameter.  This leads to poor optimization performance when attempting to learn the agent valuations.  In contrast, the likelihood for \quantalzero is continuous  and differentiable in both its behavioral parameters and the valuation.

\begin{definition}[Quantal-linear4]
    A \quantalzero (QL4) strategy for a player $i$ in a game $G$ with precision $\lambda_0>0$
    and weights $\w{max},\w{min},\w{eff},\w{fair}, \w{unif}$ is a linear sum of the form
    \[f_i(G) = \sum_{d\in\{\text{max,min,eff,fair,unif}\}} w_df_i^d(G), \]
    where the weights are constrained to lie between 0 and 1 and to sum to exactly 1.

    Each function $f$ is a soft maximization over a specific feature for each action.
    The features are: the maximum utility that $i$ can receive by playing an action; the minimum utility that $i$ can receive by playing an action; the smallest-magnitude unfairness attainable by playing an action (defined as the difference between the smallest utility and the largest; this is always negative); and the largest sum of utilities across players that is possible by playing a given action.  Formally,
    \begin{align*}
        \f{max}(G)(a_i) &= \frac{\exp[\lambda_0 \max_{a_{-i} \in A_{-i}} u_i(a_i,a_{-i})]}
                               {\sum_{a'_i \in A_i}\exp[\lambda_0 \max_{a_{-i} \in A_{-i}} u_i(a'_i,a_{-i})]} \\
        \f{min}(G)(a_i) &= \frac{\exp[\lambda_0 \min_{a_{-i} \in A_{-i}} u_i(a_i,a_{-i})]}
                               {\sum_{a'_i \in A_i}\exp[\lambda_0 \min_{a_{-i} \in A_{-i}} u_i(a'_i,a_{-i})]} \\
       \f{fair}(G)(a_i) &= \frac{\exp[\lambda_0 \max_{a_{-i} \in A_{-i}}\min_{j,j'\in N} d_{i, j}]}
                       {\sum_{a'_i\in A_i}\exp[\lambda_0 \max_{a_{-i} \in A_{-i}}\min_{j,j'\in N} d'_{i, j} ]} \\
        \noalign{where}
        d_{i, j} &= (u_j(a_i,a_{-i})-u_{j'}(a_i,a_{-i})) \\
        \noalign{and}
        d'_{i, j, A_i} &= (u_j(a'_i,a_{-i})-u_{j'}(a'_i,a_{-i})) \\
        \f{eff}(G)(a_i) &= \frac{\exp[\lambda_0 \max_{a_{-i} \in A_{-i}} \sum_{j\in N}u_j(a_i,a_{-i})]}
                               {\sum_{a'_i\in A_i}\exp[\lambda_0 \max_{a_{-i} \in A_{-i}} \sum_{j\in N}u_j(a'_i,a_{-i})]}\\
        \f{unif}(G)(a_i) &= \frac{1}{|A_i|}.
    \end{align*}
\end{definition}

\subsection{Separating Models into Strategic and Non-strategic Components}

Recalling that quantal cognitive hierarchy requires a non-strategic model in its inductive definition of behavior, it is straightforward to combine QCH and non-strategic models.  Equilibrium models such as Nash equilibrium and quantal response equilibrium can also be augmented with non-strategic behavior.  To combine equilibrium strategic models with a non-strategic component, we assume that some fraction of agents behave non-strategically, and that the strategic agents respond to this probability of non-strategic behavior as well as the behavior of the remaining probability of strategic agents.

We separate each of our models into a non-strategic component and a strategic component that responds to the non-strategic component, where each model is denoted by the naming convention "STRAT-NONSTRAT".  In this way, conventional models such as PQCH can be rethought of as PQCH-uniform, and QRE can be rethought of as QRE-none (for the sake of simplicity and in keeping with convention, we do not list the non-strategic component in a model if there is none and so QRE-none remains QRE).  
For equilibrium models we augment with a non-strategic component, we assign the parameter $\beta \in (0,1)$, to the probability of agents behaving non-strategically, with the strategic agents being assigned the remaining probability  $1-\beta$.
Unlike QCH, in which agents have heterogeneous and incorrect beliefs about the strategies of the other agents, the strategic agents in our equilibrium models augmented in this way are assumed to have correct beliefs.

The parameters for each model are thus $\theta = (\theta_{S}, \theta_{NS})$, according to Table \ref{tab:models_compared }.




\subsection{Estimation Methods}\label{section:estimation_methods}

To obtain our parameter estimate $\hat{\theta}$ of $\theta$ and $\hat{v}$ of $v$, we performed log-likelihood maximization with respect to $\hat{v}$ and $\hat{\theta}$ jointly using L-BFGS-B \cite{byrd1995limited,zhu1997algorithm}.  To evaluate the performance of our value estimation, we take the estimated value $\hat{v}$  at the maximum likelihood estimate of each model and compare it to the endowed value.  

\subsubsection{Estimation of Equilibrium Models}

To estimate equilibrium models (QRE, \qreplus, and Nash),
we chose the $(v, \lambda, \beta, \lambda_{0}, w_{L0})$ that maximized the likelihood of the
empirical behavior of the participants under the assumption that each strategic agent was quantally best responding to the empirically observed distribution $s^g_{-i}$ defined by
\[s^g_{-i}(a) = \frac{\left\lvert\{j\ne i \mid g \in G(j) \land a^g_j=a\}\right\rvert}{
\left\lvert\{j \ne i \mid g \in G(j)\}\right\rvert}. \]


For equilibrium models we maximize the following likelihood:
\begin{multline}
    \log\L(\lambda, v, \beta, \lambda_{0}, w_{L0})  = \\
    \sum_i\sum_{g\in G(i)}\log\left[ \right.
  \beta \flzero{}(G_g(v); \lambda_{0}, w_{L0})(a^g_i) \\
  +(1-\beta)QBR_i(s^g_{-i} \mid G_g(v),\lambda)(a^g_i)\left. \right]
\label{eq:eqm}
\end{multline}


where $v$ is the value parameter being estimated; $\lambda$ is the behavioral precision parameter; $\beta$ is the proportion of non-strategic agents between 0 and 1; $(\lambda_{0}, w_{L0})$ are the behavioral precision and weight parameters for \quantalzero, respectively; $G_g(v)$ is the payoff game induced from an allocation game $g$ and valuation $v$; $\hat{s}^g_{-i}$ is the empirical distribution of play in allocation game $g$; and $a^g_i$ is the action taken by participant $i$ in game $g$.%
\footnote{For models with uniform randomization as the non-strategic component, we do not estimate $(\lambda_{0}, w_{L0})$.}%
\footnote{We fix $\beta = 0$ when estimating models without a non-strategic component}

The econometric approach of computing QRE by assuming all agents are quantally responding against other agents in the empirically observed distribution of actions is commonly used \cite[e.g.][]{chen2012modeling, goeree2002quantal, bajari2005structural, noti2021behavioral}. What is not common is the simultaneous estimation of both the precision of agents $\lambda$ as well as the value parameter $v$. The previously listed works all do a two-step estimation method of either first estimating $v$ and then $\lambda \mid v$ or vice versa. This is because given an observed action $s_i(a_i)$ generated from a logit model which takes as an input observed utility  $u_i$ of the form $u_i = \lambda v$,  there are infinitely many combinations of $\lambda$ and value that could result in the same observed utility. This motivates the inclusion of a payment profile $p$ in our allocation games. Including a static payoff $p$ allows us to simultaneously estimate both $v$ and $\lambda$ by anchoring $\lambda$ to a specific scale; indeed we find that when constraining $p = 0$, our estimates are incorrect by up to an order of magnitude (refer to  Table \ref{tab:nodollar_estimates }  in the appendix).

Nash equilibrium does not have model parameters to estimate.  When estimating values using the Nash equilibrium model, we approximate best response using quantal best response with a high value of $\lambda$,\footnote{We used $\lambda=100$ in our experiments, as we found that both the predictive performance and value estimate converge at precision $\lambda \ge 100$; refer to Figure~\ref{fig:nash_summary} in the appendix for details.} and select the value that maximizes equation \eqref{eq:eqm}.
This approach allows us to select a single value that is \emph{most} consistent with best response, rather than a set of values that are consistent with all agents' best-responding.
More critically, it also ensures that every possible action has positive probability.  When assuming best response with no error model, a single action by a single agent that is not consistent with best response can lead to the entire dataset's having probability 0.  Under our approach, actions inconsistent with best response will instead be assigned a very low, but positive, probability.

\subsubsection{Estimation of Poisson Quantal Cognitive Hierarchy}

For Poisson quantal cognitive hierarchy, we estimate $(v, \lambda,\tau, \lambda_{0}, w_{L0})$ by maximizing the following likelihood:
    \begin{multline}
        \log\L  (\lambda, v, \tau, \lambda_{0}, w_{L0}) =\\ 
        \sum_i\sum_{g\in G(i)}\log\left[\right.
        L_{\tau; 0} \flzero{}(G_g(v); \lambda_{0}, w_{L0})(a^g_i) \\
        + \sum_{\ell=1}^3 L_{\tau; \ell}
        QBR_i(G_g(v),\lambda \mid \ell_{i|0:\ell-1})(a^g_i)\left. \right]
    \end{multline}
    
In contrast to the equilibrium models, the likelihood for PQCH does not treat the empirically observed distribution  as the distribution of actions being responded to; we instead find the mean parameter $\tau$ that generates a distribution which maximizes the likelihood against the empirical data.
Assuming that strategic QCH agents respond to the empirical distribution of lower-level agents would require us to estimate the levels (or posterior level distributions) for each agent, in order to estimate which agents' empirical behavior is being responded to; e.g., to determine what empirical distribution is being responded to by level-2 agents, we must first determine which agents are level-0 and level-1.  This is a much more complex estimation problem, both statistically and computationally.
For this reason, we take the simpler approach of estimating the mean parameter $\tau$ instead.\footnote{We also estimated our equilibrium models by finding the optimal parameter $\lambda$ that maximizes the likelihood against our data; we did not find a significant difference in the estimated $v$, which provides assurances that estimating against the empirical distribution provides a reasonable approximation while being much simpler to compute.  Refer to Table \ref{tab:qre_fixed_comparison} in the appendix. }
 

\subsubsection{Utilization of Panel Structure in Estimation of Values}
In our experimental setup, we collected panel data where the individual actions of each player for each game are recorded, in contrast to  other common data sets in which the actions of all agents are pooled together.  This panel structure allows the estimation of model parameters that are heterogeneous across agents but stable for a given agent $i$.
The level of an individual agent in QCH-based models is an example of a parameter that could match this description.\footnote{This same discussion applies to \qreplus, if we treat strategic agents as having a level $\ell_i=1$ and non-strategic agents as having a level $\ell_i=0$.}
\footnote{Our definition implies that every non-strategic agent plays a mixture over a number of \level{0} decision rules. However, one could also imagine a definition in which there is a population of non-strategic agents, each using a single \level{0} rule.  Under this assumption, the assignment of decision rules to agents could also fit this description of a heterogeneous but stable behavioral parameter.}
If a player's level is the same in every game, then using a likelihood that explicitly encodes this has the potential to provide more accurate estimates than one that assumes that each player's level is re-sampled before every action.  \eqref{eq:panel-L} gives the likelihood for a model with parameters $\theta$, a stable  level $\ell_i$ for agent $i$ distributed according to $\Pr(\ell_i = \ell \mid \theta)$, in which agent $i$ takes action $a$ in a game $g$ with probability $\Pr(a_i^g \mid \ell_i = \ell, \theta)$.
\begin{align}
    \log\Pr(D \mid \theta,v) &= \sum_i\sum_{g\in G(i)}\log\left[\sum_\ell \Pr(\ell_i=\ell \mid \theta) \Pr\left(a_i^{G_g(v)} \mid \ell_i = \ell, \theta\right)\right]
    \label{eq:panel-L}
\end{align}

In contrast, the likelihood for an otherwise-identical model in which each agent's level can vary between games is given by \eqref{eq:pooled-L}.
\begin{align}
    \log\Pr(D \mid \theta,v) &= \sum_i\log\left[\sum_\ell \Pr(\ell_i=\ell \mid \theta)\prod_{g\in G(i)}\Pr\left(a_i^{G_g(v)} \Mid \ell_i = \ell, \theta\right)\right] 
    \label{eq:pooled-L}
\end{align}

When running our analysis on synthetic data we find that the likelihood of \eqref{eq:pooled-L} is  more numerically stable than that of \eqref{eq:panel-L}, while returning a similar value estimate. We therefore report the parameters estimated using \eqref{eq:pooled-L} in this paper.
Our dataset is available for future research questions or models that require panel data.

\subsection{Game Construction}\label{sec:game_construction}

In our setting, $n$ subjects play a set of bimatrix payoff games $\mathcal{G}$.
To simulate our subjects playing a set of allocation games $\mathcal{A}$, we map each payoff game the participant plays to a corresponding allocation game based on an endowed valuation \truevalue of our choosing.
To transform from the payoff games presented to our subjects to the desired allocation games, we construct an allocation game in the following way:


We first select an endowed value $v^*$ that is hidden from the models we evaluate.
Then, for every cell in every payoff game $G\in\mathcal{G}$:
\begin{enumerate}
    \item We sample an allocation from a uniform distribution bounded between $0$ and $ \max(u(G))  /  v^*$.
    \item We add a payment $p \in \mathbb{R}$  such that the payoffs for each player and action match that in the original payoff game. Note that in our setup, payments can be negative.  
\end{enumerate}

This setup resolves the aforementioned issues with presenting allocation games to participants directly. Since only the utilities are presented to participants, we are able to abstract away from other effects such as arithmetic errors that might arise from having participants play the actual allocation game. This setup also allows us to specify an infinite number of games for any known \truevalue, which is useful because it allows reuse of the same dataset for multiple sets of allocation games,
as well as providing a ground truth value with which to evaluate model performance.
A more detailed pseudocode explanation can be found in Section~\ref{apx:random_allocation} of the appendix.


\subsection{Experimental Details}
We tested our approach on experimental data collected from participants on Amazon Mechanical Turk (MTurk).
We presented participants with a set of 24 $3\times3$ symmetric normal form games (the payoff games) in which each participant played against the actions of the previous participant.
All participants had at least $\>95\%$ HIT approval and had completed at least 100 HITs. 
We removed all data from participants who completed the HIT in fewer than 120 seconds, or 5 seconds per game, as there was a high correlation between participants who did this and responses at the end of the survey that were either left empty or spurious.\footnote{For example, two participants had the exact same input in the feedback field, seemingly referring to a task in an entirely different HIT.}

The payoff games were generated by randomly sampling payoffs from a uniform distribution on $[0, 100]$, and were played by participants in a randomized order.
We collected additional treatments, analyzed in Section~\ref{section:treatments} of the appendix, in which the order and type of games were varied.
Our qualitative results in these additional treatments were unchanged from those of the main treatment.

Participants were paid \$1.50 for completing the HIT, as well as a performance bonus based on their total payoffs in the games.  The  performance bonus was calculated by multiplying the payoffs achieved by the participant by  \$0.02 (USD), with the goal being to have all participants achieve an equivalent wage of at least \$10 per hour between the bonus and base payment if they had uniformly randomized and taken the maximum allowed time of 30 minutes.  

\section{Results}
Our evaluation finds that models with a rich non-strategic component perform better in value estimation from behavior in initial play than those without a non-strategic component (e.g., Nash, QRE) or those that assume that non-strategic agents uniformly randomize.  Additionally, we find that models that include a strategic component perform better at value estimation than those that assume that all agents are non-strategic, but the choice of strategic model is not as important as the rich non-strategic component in a given model.

\begin{figure}[h]
    \centering
    \includegraphics[width=0.99\linewidth]{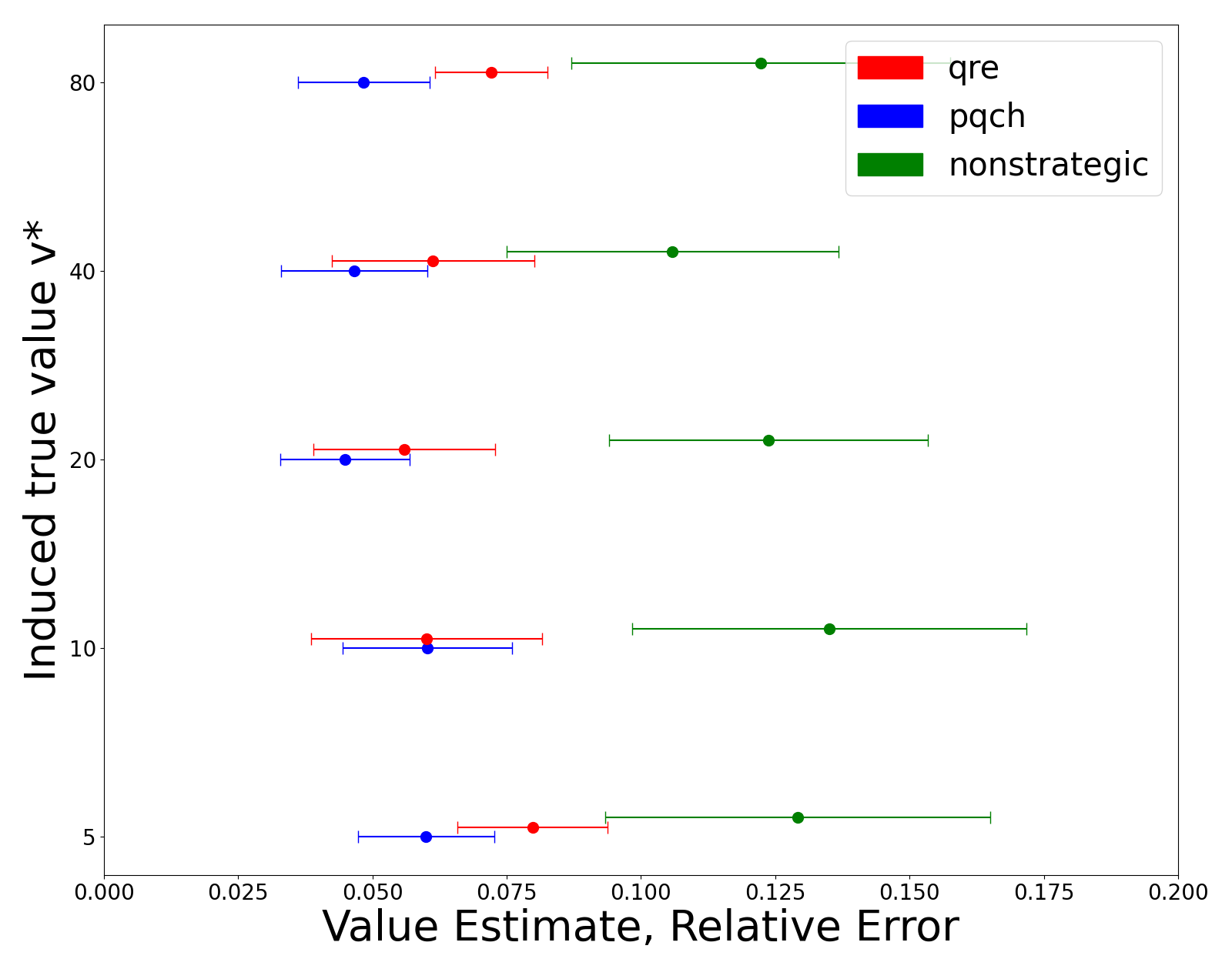}
    \caption{
    Summary plot showing values of \truevalue vs. the relative error for models with \quantalzero as the non-strategic componen The two left-most points for all values of \truevalue are models containing both a strategic component and \quantalzero as the non-strategic component; the green point indicates a model that assumes all agents are non-strategic in a  \quantalzero manner with no strategic behavior.}
    \label{fig:summary_plot}
\end{figure}

\subsection{Evaluation}

Our evaluation considered traditional equilibrium models  with no non-strategic agents (Nash, QRE), and  QCH and \qreplus where \level{0} agents were either uniform randomizers or \quantalzero agents.  We evaluated each model across multiple scenarios given \truevalue in $\mathcal{V} = [5, 10, 20, 40, 80]$.
 For each \truevalue, we generated $k = 25$ scenarios where we mapped our payoff games to a set of allocation games $\mathcal{A}_{\mathcal{G}}$ given \truevalue.
We measured each model's value estimation for each scenario using relative error, $\frac{\lvert \hat{v} - \truevalue \rvert}{\truevalue}$. We chose to normalize the error to account for the differing scale of values in $\mathcal{V}$.
The value estimate for each scenario was evaluated using using the mean value estimate of 10 rounds of 10-fold cross-validation, with the test set being used to evaluate behavioral prediction.
The mean value estimate for each scenario are distributed according to a Student's $t$-distribution \cite[e.g.][]{witten2002data}. We say that a model outperforms another in value estimation when the 95\% confidence intervals do not overlap.

Figure~\ref{fig:summary_plot} and Table~\ref{tab:estimate_breakdown} show the performance in value estimation across models, with Figure~\ref{fig:summary_plot} being a visualization of the data in Table~\ref{tab:estimate_breakdown}.
Behavioral models with \quantalzero as the non-strategic component outperform classical equilibrium models in terms of value estimation across every endowed value \truevalue that we evaluated.
We find that using \quantalzero as the non-strategic component outperforms corresponding strategic models which use a uniform non-strategic component, regardless of the choice of strategic~model.  This leads us to conclude that modelling non-strategic behavior is more important than the choice of strategic model.  Note, however, that None-QL4 does not perform as well as PQCH-QL4 or QRE-QL4, which suggests that a strategic component in the model is still necessary.  Another observation is that models containing QL4 remain stable across values of \truevalue; the mean relative error for QL4 models  varies at most by 2\%,  in contrast to classic equilibrium models or uniform non-strategic augmented models in which the relative errors differ by an order of magnitude from each other depending on \truevalue . This leads us to conclude that QL4 leads to a more \emph{reliable} estimate of values.

We demonstrate the importance of obtaining accurate value estimates in Table~\ref{tab:average_welfare_mse}. We first obtain an estimate of $\theta$ and $v$ on half of the games in our dataset ($m = 12$). Using the estimated value $\hat{v}$ and behavioral parameters $\hat{\theta}$, we then predict the average subject welfare for the remaining half of games that were held out.  We then compute the relative error of the  predicted welfare against the empirically observed average welfare of subjects.  This evaluation requires a model to be accurate in both its estimation of behavioral parameters as well as that of values; a model with  an accurate value estimate but a poor prediction of behavior would perform poorly, and vice versa. PQCH-QL4 and QRE-QL4 once again perform the best at this task, with Nash being noticeably poor at welfare prediction, especially at lower values of $v$.  This pattern persists across models;  welfare estimates are worse for lower values of \truevalue compared to higher values, albeit at a much larger scale for Nash and for models with a uniform non-strategic component. The final note here is that QRE-None outperforms QRE-uniform across the board, which shows that an arbitrary level-0 model is not sufficient to improve performance, but a rich level-0 model is required.

\begin{table*}[!htp]\centering
\caption{Relative error by \truevalue, with confidence interval in parentheses. Bold cells indicate best performing model for each \truevalue. Italicized cells indicate models which are not significantly different from the best performing one. QRE-QL4 indicates a model in which a fraction of agents are behaving non-strategically in a QL4 manner while the remaining agents are in QRE with themselves and non-strategic agents. None-QL4 indicates a model in which all agents are behaving non-strategically. }\label{tab:estimate_breakdown}
\begin{tabular}{ccccccc}\toprule
        \multicolumn{2}{c}{Component} & \multicolumn{5}{c}{$\truevalue$} \\
            \cmidrule(lr){1-2}\cmidrule{3-7}
Strategic & Non-strategic &5 & 10 &20 &40 &80 \\\midrule
Nash &none &10.41, (8.01, 12.8) &2.88, (2.06, 3.71) &0.64, (0.44, 0.83) &0.29, (0.18, 0.4) &0.2, (0.15, 0.25) \\
QRE &none &0.14, (0.1, 0.18) &0.11, (0.08, 0.14) &0.11, (0.07, 0.14) &0.13, (0.09, 0.17) &0.1, (0.07, 0.12) \\
QRE & uniform & 8.27, (5.14, 11.4) & 2.04, (1.21, 2.87) &	0.37, (0.18, 0.56) &	0.13, (0.09, 0.17) &	0.1, (0.07, 0.13) \\
PQCH &uniform &1.93, (0.68, 3.19) &0.32, (0.16, 0.48) &0.12, (0.08, 0.16) &0.09, (0.07, 0.12) &\textit{0.08, (0.05, 0.11)}\\
PQCH &QL4 &\textbf{0.06, (0.05, 0.07)} &\textbf{0.06, (0.04, 0.08)} &\textbf{0.05, (0.03, 0.06)} &\textbf{0.05, (0.03, 0.06)} &\textbf{0.05, (0.04, 0.06)} \\
QRE &QL4 &\textit{0.08, (0.07, 0.09)} & \textit{0.06, (0.04, 0.08)} & \textit{0.06, (0.04, 0.07)} &\textit{0.06, (0.04, 0.08)} &0.07, (0.06, 0.08) \\
none &QL4 &0.13, (0.09, 0.17) &0.14, (0.1, 0.17) &0.12, (0.09, 0.15) &0.11, (0.08, 0.14) &0.12, (0.09, 0.16) \\
\bottomrule
\end{tabular}
\end{table*}

\begin{table*}[!htp]\centering
\caption{Relative error of predicted average per game welfare by \truevalue. In each scenario,  the estimated valuation and model parameters $\theta$ from half the games are used to predict the average game welfare per subject on the other half and is compared against the empirically observed average welfare. Bold cells indicate models with the lowest MSE.}\label{tab:average_welfare_mse}
\begin{tabular}{ccccccc}\toprule
            \multicolumn{2}{c}{Component} & \multicolumn{5}{c}{$\truevalue$} \\
            \cmidrule(lr){1-2}\cmidrule{3-7}
Strategic &Non-strategic &5 &10 &20 &40 &80 \\\midrule
Nash &none &16.81 &5.42 &1.75 &0.41 &0.30 \\
QRE &none &0.24 &0.20 &0.20 &0.18 &0.12 \\
QRE &uniform &12.96 &4.26 &0.55 &0.16 &0.15 \\
PQCH-uniform &uniform &5.86 &0.52 &0.22 &0.13 &0.11 \\
PQCH-QL4 &QL4 &\textbf{0.12} &0.12 &\textbf{0.09} &\textbf{0.09} &\textbf{0.07} \\
QRE-QL4 &QL4 &\textbf{0.12} &\textbf{0.11} &\textbf{0.09} &0.10 &0.08 \\
none &QL4 &0.56 &0.27 &0.39 &0.18 &0.18 \\
\bottomrule
\end{tabular}
\end{table*}


\subsection{Contribution of Strategic vs. Non-strategic Components of the Model}
\label{section:ql4_contribution}
We attempt to quantify the contribution of \quantalzero to the observed improvement in value estimation.
We compare the cross-product of our strategic and non-strategic components as discussed in Section~\ref{section:behavioral_models} and find that QL4 outperforms any of the other non-strategic models considered.
In addition to comparing \quantalzero and uniformly randomization,
in this section we include \linear4 from \citep{wright2017predicting} as well as a differentiable version of \linear4 we refer to as differentiable-\linear4 (DL4) where $\lambda_{0} = 1$, which gives us a differentiable function with respect to $v$ without adding an additional degree of freedom.  

For each resulting model resulting from the cross-product of strategic and non-strategic components,  we take each of our scenarios for each value \truevalue (n = 125) and report the percentage of the time that the relative error of $\hat{v}$ falls below a threshold $\alpha$ (i.e., the error falls within 10\% accuracy).
We sampled 1000 bootstrapped samples from our empirically observed data $\mathcal{D}$ and did this for each bootstrapped sample, reporting the median percentage each model falls within our threshold with the lower and upper bounds being the middle  95\% of the bootstrapped estimates as outlined in \citep{cohen1995empirical}.  
Doing this allows us to see how well a given non-strategic component performs at recovering \truevalue, regardless of the strategic component being used in the model.  The results demonstrate the advantages of \quantalzero, as it performs strictly better than uniform and \linear4, and outperforms differentiable-\linear4, although not significantly.
The results of this test are reported in Table \ref{tab:component-matrix }.

There are two reasons why \linear4 performs poorly as a non-strategic model: the first is that as a non-continuous function of $v$, it is not differentiable with respect to $v$ and so our optimization procedure fails to reliably find the value that maximizes likelihood;
checks on synthetic data show that the likelihood returned by the estimator is often worse than the likelihood at the known ground truth value. A second possible reason is due to the lack of quantal response in non-strategic agents; if we believe that strategic agents quantally respond to their payoffs, it stands to reason that  non-strategic agents do so as well. This provides a possible explanation for why \quantalzero outperforms differentiable-linear4.




\begin{table*}[!htp]\centering
\caption{Percentage of the time that a model's relative error falls within 10\% of \truevalue, across all values in  $\mathcal{V}$.  Each cell corresponds to a model STRAT-NONSTRAT with the row indicating the strategic model and column indicating the non-strategic one.  QL4 (rightmost column) outperforms all other non-strategic components regardless of the strategic model. Here, the confidence intervals are derived from a $k$ bootstrapped samples of the observed data, with $k= 1000$. Cells marked "n/a" do not have a conceivable model that elicits an estimate of values. Cells containing 0 mean that none of the bootstrapped samples had a value estimate that fell within 10\% of \truevalue. }\label{tab:component-matrix }
\begin{tabular}{ccccccc}\toprule
        \multicolumn{1}{c}{Strategic Component} & \multicolumn{5}{c}{Non-strategic Component} \\
            \cmidrule(lr){1-1}\cmidrule{2-6}
&None &Uniform &L4 &DL4 &QL4 \\\midrule
Nash &0.1200 (0.0640 0.1760) &0.0960 (0.0480 0.1520) & 0.00 &0.0480 (0.0160 0.0880) &0.6880 (0.5040 0.8400) \\
QRE &0.4960 (0.4080 0.5680) &0.3280 (0.2560 0.4080) & 0.00 &0.7120 (0.6160 0.8160) &0.7040 (0.6000 0.8160) \\
PQCH &n/a &0.5200 (0.4160 0.6400) & 0.00 &0.6240 (0.4960 0.7440) &0.8440 (0.7200 0.9360) \\
None &n/a &n/a & 0.00 &0.00 &0.4480 (0.3040 0.5840) \\
\bottomrule
\end{tabular}
\end{table*}

\section{Conclusion}
This paper examines the benefit of using behavioral models for value estimation.
%
Behavioral models typically include parameters that must be estimated from the data.
Using a novel experimental design, we demonstrate that estimating these behavioral parameters simultaneously with value parameters is feasible, and leads to more reliably accurate value estimations from initial play than value estimates based on the standard strong equilibrium assumption.

We introduce a new specification of \level{0} behavior called \emph{\quantalzero}, and a new behavioral model called \emph{\qreplus} that extends quantal response equilibrium to settings that contain non-strategic agents, who are responsive to their own preferences but do not reason about other agents.
Our results show that models that include a rich \level{0} specification perform better at estimating values from initial play.  These results strongly argue for the importance of explicitly modeling non-strategic behavior rather than treating it as noise, especially in contexts such as initial play in which equilibrium is unlikely to have been reached.

\subsection{Future Work}


There are a number of directions in which this work could be extended. We made a simplifying assumption that all agents shared a homogeneous value, but an important future direction would be estimating heterogeneous values. Further to this direction, we could extend this work to estimating individual behavioral parameters of agents.  This would allow us to model differences in behavior based on heterogeneous beliefs about the values of others, which could lead to better individual welfare.  As mentioned in Section \ref{section:estimation_methods}, we could also extend the approach of responding to empirically observed distributions in models of iterated reasoning, which would allow us to move beyond specifying a distribution over levels.

Using our framework of separating models into a strategic and non-strategic component, we could examine other models of non-strategic behavior (for example, using the model from \citet{fudenberg2019predicting} as a non-strategic model) beyond those discussed in this paper.
Finally, further extending non-strategic behavioral models is an important direction for future work.
This could take the form of extending QL4 to be more predictive, or evaluating domain-specific models of nonstrategic behavior.

\newpage
\section*{Acknowledgements}
We would like to acknowledge Michalis Mamakos for his participation in an earlier iteration of this project. We would also like to acknowledge the turkers who participated in our experiment.
 This work was funded in part by NSF award CCF-1934931.
The third author holds a Canada CIFAR AI Chair through the Alberta Machine Intelligence Institute. 

\bibliographystyle{aamas-ACM-Reference-Format}
\bibliography{ref}


\begin{thebibliography}{30}


\ifx \showCODEN    \undefined \def \showCODEN     #1{\unskip}     \fi
\ifx \showDOI      \undefined \def \showDOI       #1{#1}\fi
\ifx \showISBNx    \undefined \def \showISBNx     #1{\unskip}     \fi
\ifx \showISBNxiii \undefined \def \showISBNxiii  #1{\unskip}     \fi
\ifx \showISSN     \undefined \def \showISSN      #1{\unskip}     \fi
\ifx \showLCCN     \undefined \def \showLCCN      #1{\unskip}     \fi
\ifx \shownote     \undefined \def \shownote      #1{#1}          \fi
\ifx \showarticletitle \undefined \def \showarticletitle #1{#1}   \fi
\ifx \showURL      \undefined \def \showURL       {\relax}        \fi
\providecommand\bibfield[2]{#2}
\providecommand\bibinfo[2]{#2}
\providecommand\natexlab[1]{#1}
\providecommand\showeprint[2][]{arXiv:#2}

\bibitem[\protect\citeauthoryear{Agarwal and Somaini}{Agarwal and
  Somaini}{2018}]%
        {agarwal2018demand}
\bibfield{author}{\bibinfo{person}{Nikhil Agarwal} {and} \bibinfo{person}{Paulo
  Somaini}.} \bibinfo{year}{2018}\natexlab{}.
\newblock \showarticletitle{Demand analysis using strategic reports: An
  application to a school choice mechanism}.
\newblock \bibinfo{journal}{\emph{Econometrica}} \bibinfo{volume}{86},
  \bibinfo{number}{2} (\bibinfo{year}{2018}), \bibinfo{pages}{391--444}.
\newblock


\bibitem[\protect\citeauthoryear{Athey and Haile}{Athey and Haile}{2007}]%
        {athey2007nonparametric}
\bibfield{author}{\bibinfo{person}{Susan Athey} {and} \bibinfo{person}{Philip~A
  Haile}.} \bibinfo{year}{2007}\natexlab{}.
\newblock \showarticletitle{Nonparametric approaches to auctions}.
\newblock \bibinfo{journal}{\emph{Handbook of econometrics}}
  \bibinfo{volume}{6} (\bibinfo{year}{2007}), \bibinfo{pages}{3847--3965}.
\newblock


\bibitem[\protect\citeauthoryear{Athey and Nekipelov}{Athey and
  Nekipelov}{2010}]%
        {athey2010structural}
\bibfield{author}{\bibinfo{person}{Susan Athey} {and} \bibinfo{person}{Denis
  Nekipelov}.} \bibinfo{year}{2010}\natexlab{}.
\newblock \showarticletitle{A structural model of sponsored search advertising
  auctions}. In \bibinfo{booktitle}{\emph{Sixth ad auctions workshop}},
  Vol.~\bibinfo{volume}{15}.
\newblock


\bibitem[\protect\citeauthoryear{Bajari and Hortacsu}{Bajari and
  Hortacsu}{2005}]%
        {bajari2005structural}
\bibfield{author}{\bibinfo{person}{Patrick Bajari} {and} \bibinfo{person}{Ali
  Hortacsu}.} \bibinfo{year}{2005}\natexlab{}.
\newblock \showarticletitle{Are structural estimates of auction models
  reasonable? Evidence from experimental data}.
\newblock \bibinfo{journal}{\emph{Journal of Political economy}}
  \bibinfo{volume}{113}, \bibinfo{number}{4} (\bibinfo{year}{2005}),
  \bibinfo{pages}{703--741}.
\newblock


\bibitem[\protect\citeauthoryear{Byrd, Lu, Nocedal, and Zhu}{Byrd
  et~al\mbox{.}}{1995}]%
        {byrd1995limited}
\bibfield{author}{\bibinfo{person}{Richard~H Byrd}, \bibinfo{person}{Peihuang
  Lu}, \bibinfo{person}{Jorge Nocedal}, {and} \bibinfo{person}{Ciyou Zhu}.}
  \bibinfo{year}{1995}\natexlab{}.
\newblock \showarticletitle{A limited memory algorithm for bound constrained
  optimization}.
\newblock \bibinfo{journal}{\emph{SIAM Journal on scientific computing}}
  \bibinfo{volume}{16}, \bibinfo{number}{5} (\bibinfo{year}{1995}),
  \bibinfo{pages}{1190--1208}.
\newblock


\bibitem[\protect\citeauthoryear{Calsamiglia, Fu, and G{\"u}ell}{Calsamiglia
  et~al\mbox{.}}{2020}]%
        {calsamiglia2020structural}
\bibfield{author}{\bibinfo{person}{Caterina Calsamiglia}, \bibinfo{person}{Chao
  Fu}, {and} \bibinfo{person}{Maia G{\"u}ell}.}
  \bibinfo{year}{2020}\natexlab{}.
\newblock \showarticletitle{Structural estimation of a model of school choices:
  The boston mechanism versus its alternatives}.
\newblock \bibinfo{journal}{\emph{Journal of Political Economy}}
  \bibinfo{volume}{128}, \bibinfo{number}{2} (\bibinfo{year}{2020}),
  \bibinfo{pages}{642--680}.
\newblock


\bibitem[\protect\citeauthoryear{Camerer, Ho, and Chong}{Camerer
  et~al\mbox{.}}{2004}]%
        {camerer2004cognitive}
\bibfield{author}{\bibinfo{person}{Colin~F Camerer}, \bibinfo{person}{Teck-Hua
  Ho}, {and} \bibinfo{person}{Juin-Kuan Chong}.}
  \bibinfo{year}{2004}\natexlab{}.
\newblock \showarticletitle{A cognitive hierarchy model of games}.
\newblock \bibinfo{journal}{\emph{The Quarterly Journal of Economics}}
  \bibinfo{volume}{119}, \bibinfo{number}{3} (\bibinfo{year}{2004}),
  \bibinfo{pages}{861--898}.
\newblock


\bibitem[\protect\citeauthoryear{Chawla, Hartline, and Nekipelov}{Chawla
  et~al\mbox{.}}{2016}]%
        {CHN-16}
\bibfield{author}{\bibinfo{person}{Shuchi Chawla}, \bibinfo{person}{Jason
  Hartline}, {and} \bibinfo{person}{Denis Nekipelov}.}
  \bibinfo{year}{2016}\natexlab{}.
\newblock \showarticletitle{A/B testing of auctions}. In
  \bibinfo{booktitle}{\emph{Proceedings of the 2016 ACM Conference on Economics
  and Computation}}. \bibinfo{pages}{19--20}.
\newblock


\bibitem[\protect\citeauthoryear{Chen, Su, and Zhao}{Chen
  et~al\mbox{.}}{2012}]%
        {chen2012modeling}
\bibfield{author}{\bibinfo{person}{Yefen Chen}, \bibinfo{person}{Xuanming Su},
  {and} \bibinfo{person}{Xiaobo Zhao}.} \bibinfo{year}{2012}\natexlab{}.
\newblock \showarticletitle{Modeling bounded rationality in capacity allocation
  games with the quantal response equilibrium}.
\newblock \bibinfo{journal}{\emph{Management Science}} \bibinfo{volume}{58},
  \bibinfo{number}{10} (\bibinfo{year}{2012}), \bibinfo{pages}{1952--1962}.
\newblock


\bibitem[\protect\citeauthoryear{Cohen}{Cohen}{1995}]%
        {cohen1995empirical}
\bibfield{author}{\bibinfo{person}{Paul~R Cohen}.}
  \bibinfo{year}{1995}\natexlab{}.
\newblock \bibinfo{booktitle}{\emph{Empirical methods for artificial
  intelligence}}. Vol.~\bibinfo{volume}{139}.
\newblock \bibinfo{publisher}{MIT press Cambridge}.
\newblock


\bibitem[\protect\citeauthoryear{Crawford and Iriberri}{Crawford and
  Iriberri}{2007}]%
        {crawford2007level}
\bibfield{author}{\bibinfo{person}{Vincent~P Crawford} {and}
  \bibinfo{person}{Nagore Iriberri}.} \bibinfo{year}{2007}\natexlab{}.
\newblock \showarticletitle{Level-k auctions: Can a nonequilibrium model of
  strategic thinking explain the winner's curse and overbidding in
  private-value auctions?}
\newblock \bibinfo{journal}{\emph{Econometrica}} \bibinfo{volume}{75},
  \bibinfo{number}{6} (\bibinfo{year}{2007}), \bibinfo{pages}{1721--1770}.
\newblock


\bibitem[\protect\citeauthoryear{Fudenberg, Gao, and Liang}{Fudenberg
  et~al\mbox{.}}{2020}]%
        {fudenberg2020flexible}
\bibfield{author}{\bibinfo{person}{Drew Fudenberg}, \bibinfo{person}{Wayne
  Gao}, {and} \bibinfo{person}{Annie Liang}.} \bibinfo{year}{2020}\natexlab{}.
\newblock \showarticletitle{How Flexible is that Functional Form? Quantifying
  the Restrictiveness of Theories}.
\newblock \bibinfo{journal}{\emph{arXiv preprint arXiv:2007.09213}}
  (\bibinfo{year}{2020}).
\newblock


\bibitem[\protect\citeauthoryear{Fudenberg, Kleinberg, Liang, and
  Mullainathan}{Fudenberg et~al\mbox{.}}{2019}]%
        {fudenberg2019measuring}
\bibfield{author}{\bibinfo{person}{Drew Fudenberg}, \bibinfo{person}{Jon
  Kleinberg}, \bibinfo{person}{Annie Liang}, {and} \bibinfo{person}{Sendhil
  Mullainathan}.} \bibinfo{year}{2019}\natexlab{}.
\newblock \showarticletitle{Measuring the completeness of theories}.
\newblock  (\bibinfo{year}{2019}).
\newblock


\bibitem[\protect\citeauthoryear{Fudenberg and Liang}{Fudenberg and
  Liang}{2019}]%
        {fudenberg2019predicting}
\bibfield{author}{\bibinfo{person}{Drew Fudenberg} {and} \bibinfo{person}{Annie
  Liang}.} \bibinfo{year}{2019}\natexlab{}.
\newblock \showarticletitle{Predicting and understanding initial play}.
\newblock \bibinfo{journal}{\emph{American Economic Review}}
  \bibinfo{volume}{109}, \bibinfo{number}{12} (\bibinfo{year}{2019}),
  \bibinfo{pages}{4112--41}.
\newblock


\bibitem[\protect\citeauthoryear{Goeree, Holt, and Palfrey}{Goeree
  et~al\mbox{.}}{2002}]%
        {goeree2002quantal}
\bibfield{author}{\bibinfo{person}{Jacob~K Goeree}, \bibinfo{person}{Charles~A
  Holt}, {and} \bibinfo{person}{Thomas~R Palfrey}.}
  \bibinfo{year}{2002}\natexlab{}.
\newblock \showarticletitle{Quantal response equilibrium and overbidding in
  private-value auctions}.
\newblock \bibinfo{journal}{\emph{Journal of Economic Theory}}
  \bibinfo{volume}{104}, \bibinfo{number}{1} (\bibinfo{year}{2002}),
  \bibinfo{pages}{247--272}.
\newblock


\bibitem[\protect\citeauthoryear{Guerre, Perrigne, and Vuong}{Guerre
  et~al\mbox{.}}{2000}]%
        {guerre2000optimal}
\bibfield{author}{\bibinfo{person}{Emmanuel Guerre}, \bibinfo{person}{Isabelle
  Perrigne}, {and} \bibinfo{person}{Quang Vuong}.}
  \bibinfo{year}{2000}\natexlab{}.
\newblock \showarticletitle{Optimal nonparametric estimation of first-price
  auctions}.
\newblock \bibinfo{journal}{\emph{Econometrica}} \bibinfo{volume}{68},
  \bibinfo{number}{3} (\bibinfo{year}{2000}), \bibinfo{pages}{525--574}.
\newblock


\bibitem[\protect\citeauthoryear{He}{He}{2015}]%
        {he2015gaming}
\bibfield{author}{\bibinfo{person}{Yinghua He}.}
  \bibinfo{year}{2015}\natexlab{}.
\newblock \showarticletitle{Gaming the Boston school choice mechanism in
  Beijing}.
\newblock  (\bibinfo{year}{2015}).
\newblock


\bibitem[\protect\citeauthoryear{Hwang}{Hwang}{2015}]%
        {hwang2015robust}
\bibfield{author}{\bibinfo{person}{Il~Myoung Hwang}.}
  \bibinfo{year}{2015}\natexlab{}.
\newblock \emph{\bibinfo{title}{A robust redesign of high school match}}.
\newblock \bibinfo{thesistype}{Ph.\,D. Dissertation}. \bibinfo{school}{The
  University of Chicago}.
\newblock


\bibitem[\protect\citeauthoryear{Ling, Fang, and Kolter}{Ling
  et~al\mbox{.}}{2018}]%
        {ling2018game}
\bibfield{author}{\bibinfo{person}{Chun~Kai Ling}, \bibinfo{person}{Fei Fang},
  {and} \bibinfo{person}{J~Zico Kolter}.} \bibinfo{year}{2018}\natexlab{}.
\newblock \showarticletitle{What game are we playing? end-to-end learning in
  normal and extensive form games}.
\newblock \bibinfo{journal}{\emph{arXiv preprint arXiv:1805.02777}}
  (\bibinfo{year}{2018}).
\newblock


\bibitem[\protect\citeauthoryear{McFadden}{McFadden}{1981}]%
        {mcfadden1981econometric}
\bibfield{author}{\bibinfo{person}{Daniel McFadden}.}
  \bibinfo{year}{1981}\natexlab{}.
\newblock \showarticletitle{Econometric models of probabilistic choice}.
\newblock \bibinfo{journal}{\emph{Structural analysis of discrete data with
  econometric applications}}  \bibinfo{volume}{198272} (\bibinfo{year}{1981}).
\newblock


\bibitem[\protect\citeauthoryear{McKelvey and Palfrey}{McKelvey and
  Palfrey}{1995}]%
        {mckelvey1995quantal}
\bibfield{author}{\bibinfo{person}{Richard~D McKelvey} {and}
  \bibinfo{person}{Thomas~R Palfrey}.} \bibinfo{year}{1995}\natexlab{}.
\newblock \showarticletitle{Quantal response equilibria for normal form games}.
\newblock \bibinfo{journal}{\emph{Games and economic behavior}}
  \bibinfo{volume}{10}, \bibinfo{number}{1} (\bibinfo{year}{1995}),
  \bibinfo{pages}{6--38}.
\newblock


\bibitem[\protect\citeauthoryear{Nekipelov, Syrgkanis, and Tardos}{Nekipelov
  et~al\mbox{.}}{2015}]%
        {nekipelov2015econometrics}
\bibfield{author}{\bibinfo{person}{Denis Nekipelov}, \bibinfo{person}{Vasilis
  Syrgkanis}, {and} \bibinfo{person}{Eva Tardos}.}
  \bibinfo{year}{2015}\natexlab{}.
\newblock \showarticletitle{Econometrics for learning agents}. In
  \bibinfo{booktitle}{\emph{Proceedings of the sixteenth acm conference on
  economics and computation}}. \bibinfo{pages}{1--18}.
\newblock


\bibitem[\protect\citeauthoryear{Noti}{Noti}{2021}]%
        {noti2021behavioral}
\bibfield{author}{\bibinfo{person}{Gali Noti}.}
  \bibinfo{year}{2021}\natexlab{}.
\newblock \showarticletitle{From Behavioral Theories to Econometrics: Inferring
  Preferences of Human Agents from Data on Repeated Interactions}. In
  \bibinfo{booktitle}{\emph{Proceedings of the AAAI Conference on Artificial
  Intelligence}}, Vol.~\bibinfo{volume}{35}. \bibinfo{pages}{5637--5646}.
\newblock


\bibitem[\protect\citeauthoryear{Paarsch, Hong, et~al\mbox{.}}{Paarsch
  et~al\mbox{.}}{2006}]%
        {paarsch2006introduction}
\bibfield{author}{\bibinfo{person}{Harry~J Paarsch}, \bibinfo{person}{Han
  Hong}, {et~al\mbox{.}}} \bibinfo{year}{2006}\natexlab{}.
\newblock \showarticletitle{An introduction to the structural econometrics of
  auction data}.
\newblock \bibinfo{journal}{\emph{MIT Press Books}}  \bibinfo{volume}{1}
  (\bibinfo{year}{2006}).
\newblock


\bibitem[\protect\citeauthoryear{Stahl and Wilson}{Stahl and Wilson}{1994}]%
        {stahl1994experimental}
\bibfield{author}{\bibinfo{person}{Dale~O Stahl} {and} \bibinfo{person}{Paul~W
  Wilson}.} \bibinfo{year}{1994}\natexlab{}.
\newblock \showarticletitle{Experimental evidence on players' models of other
  players}.
\newblock \bibinfo{journal}{\emph{Journal of economic behavior \&
  organization}} \bibinfo{volume}{25}, \bibinfo{number}{3}
  (\bibinfo{year}{1994}), \bibinfo{pages}{309--327}.
\newblock


\bibitem[\protect\citeauthoryear{Witten and Frank}{Witten and Frank}{2002}]%
        {witten2002data}
\bibfield{author}{\bibinfo{person}{Ian~H Witten} {and} \bibinfo{person}{Eibe
  Frank}.} \bibinfo{year}{2002}\natexlab{}.
\newblock \showarticletitle{Data mining: practical machine learning tools and
  techniques with Java implementations}.
\newblock \bibinfo{journal}{\emph{Acm Sigmod Record}} \bibinfo{volume}{31},
  \bibinfo{number}{1} (\bibinfo{year}{2002}), \bibinfo{pages}{76--77}.
\newblock


\bibitem[\protect\citeauthoryear{Wright and Leyton-Brown}{Wright and
  Leyton-Brown}{2017}]%
        {wright2017predicting}
\bibfield{author}{\bibinfo{person}{James~R Wright} {and} \bibinfo{person}{Kevin
  Leyton-Brown}.} \bibinfo{year}{2017}\natexlab{}.
\newblock \showarticletitle{Predicting human behavior in unrepeated,
  simultaneous-move games}.
\newblock \bibinfo{journal}{\emph{Games and Economic Behavior}}
  \bibinfo{volume}{106} (\bibinfo{year}{2017}), \bibinfo{pages}{16--37}.
\newblock


\bibitem[\protect\citeauthoryear{Wright and Leyton-Brown}{Wright and
  Leyton-Brown}{2019}]%
        {wright2019level}
\bibfield{author}{\bibinfo{person}{James~R Wright} {and} \bibinfo{person}{Kevin
  Leyton-Brown}.} \bibinfo{year}{2019}\natexlab{}.
\newblock \showarticletitle{Level-0 models for predicting human behavior in
  games}.
\newblock \bibinfo{journal}{\emph{Journal of Artificial Intelligence Research}}
   \bibinfo{volume}{64} (\bibinfo{year}{2019}), \bibinfo{pages}{357--383}.
\newblock


\bibitem[\protect\citeauthoryear{Wright and Leyton-Brown}{Wright and
  Leyton-Brown}{2022}]%
        {wright2022formal}
\bibfield{author}{\bibinfo{person}{James~R Wright} {and} \bibinfo{person}{Kevin
  Leyton-Brown}.} \bibinfo{year}{2022}\natexlab{}.
\newblock \showarticletitle{A Formal Separation Between Strategic and
  Nonstrategic Behavior}.
\newblock \bibinfo{journal}{\emph{arXiv preprint arXiv:1812.11571}}
  (\bibinfo{year}{2022}).
\newblock


\bibitem[\protect\citeauthoryear{Zhu, Byrd, Lu, and Nocedal}{Zhu
  et~al\mbox{.}}{1997}]%
        {zhu1997algorithm}
\bibfield{author}{\bibinfo{person}{Ciyou Zhu}, \bibinfo{person}{Richard~H
  Byrd}, \bibinfo{person}{Peihuang Lu}, {and} \bibinfo{person}{Jorge Nocedal}.}
  \bibinfo{year}{1997}\natexlab{}.
\newblock \showarticletitle{Algorithm 778: L-BFGS-B: Fortran subroutines for
  large-scale bound-constrained optimization}.
\newblock \bibinfo{journal}{\emph{ACM Transactions on mathematical software
  (TOMS)}} \bibinfo{volume}{23}, \bibinfo{number}{4} (\bibinfo{year}{1997}),
  \bibinfo{pages}{550--560}.
\newblock


\end{thebibliography}

\newpage
\appendix
\hbox{}\thispagestyle{empty}

\section{Appendix}
\subsection{Additional Experimental Details}
\subsubsection{Difference Between Treatments}\label{section:treatments}

When conducting the data collection process on Mturk, we varied 2 conditions for a total of 4 treatments. 
\begin{enumerate}
    \item We chose the payoff games according to two procedures.  In the first condition, we used randomly-generated payoff games with no further filtering.  In the second condition, we only used randomly generated payoff games for which no level-$k$ strategy was similar to the level-$(k-1)$ strategy, when any of the \texttt{linear4} decision rules were used as a \level{0} strategy.\footnote{Our motivation for the filtered procedure was to enable the estimation of the parameters of the cognitive hierarchy behavioral model for individual participants; however, this proved to be infeasible under realistically small values of the precision parameter~$\lambda$.}
          We refer to the games from the first condition as the unfiltered games, and the games from the second  condition as the filtered games.
    \item In the ``ordered'' condition, we showed all payoff games to the participants in the same order. In the ``randomized'' condition, we showed the payoff games to each participant in a randomized order.
\end{enumerate}

The results reported on in the main paper are  that of the nonfiltered randomized treatment.

\begin{table*}[!htp]\centering
\caption{Summary Statistics for Experimental Treatments}\label{tab:experiment-stats}
\scriptsize
\begin{tabular}{cccccc}\toprule
Treatment & \# Participants & Avg.\ Bonus  &Avg.\ Total &Avg.\ Time (minutes:seconds) \\\midrule
Filtered ordered &303 &2.50 &4.00 & 9:30 \\ 
Nonfiltered ordered &179 &2.71 &4.21 & 9:11\\ 
Filtered randomized &181 &2.38 &3.88 & 9:31 \\ 
Nonfiltered randomized &185 &2.58 &4.08 & 10:18 \\ 
\bottomrule
\end{tabular}
\end{table*}



\subsection{Experimental Details}
\begin{figure*}
\centering
\includegraphics[width=\textwidth]{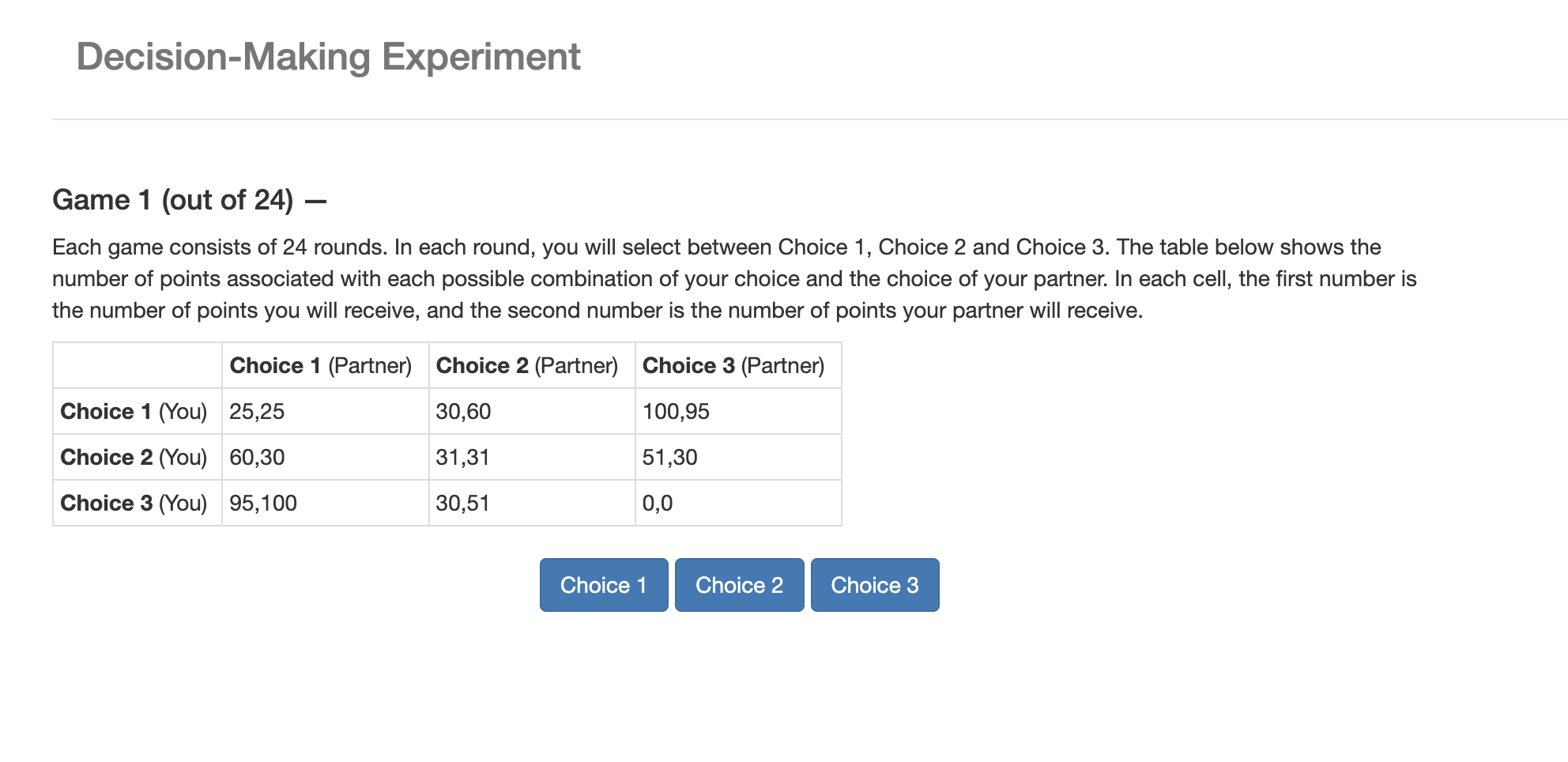}
\caption{The experiment webpage presented to MTurk Participants.}
\label{fig:experiment_interface}
\end{figure*}

\begin{figure*}
\centering
\includegraphics[width=\textwidth]{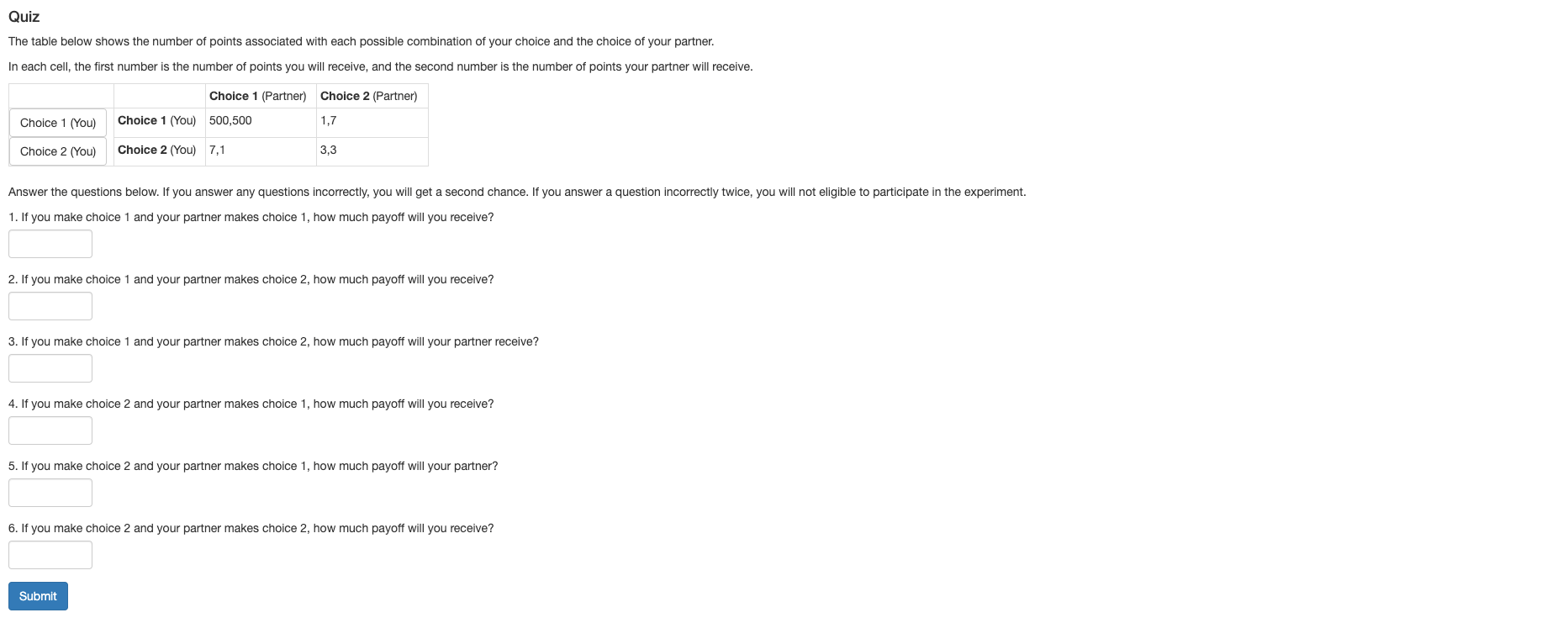}
\caption{The screening quiz presented to MTurk Participants. Participants were allowed 3 attempts on the quiz before being rejected for the HIT. }
\label{fig:quiz}
\end{figure*}

\begin{figure*}
\centering
\includegraphics[width=\textwidth]{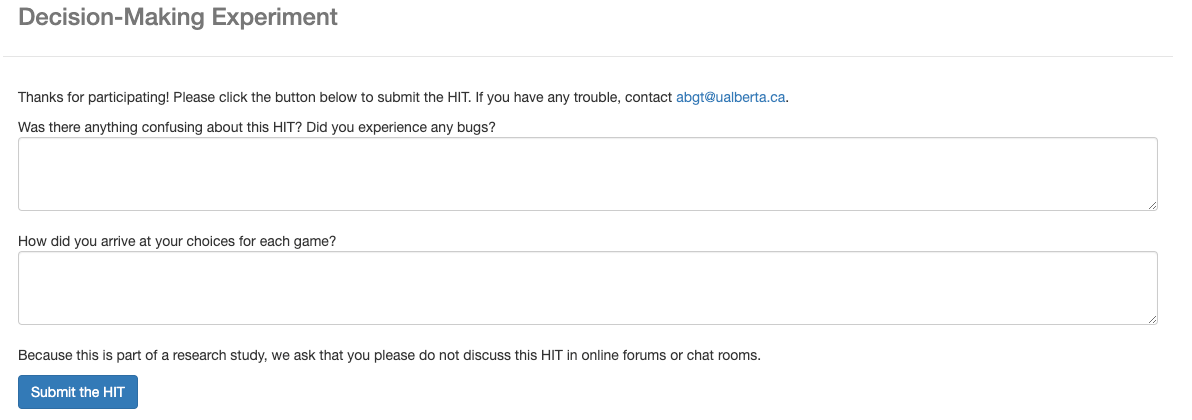}
\caption{The exit survey presented to MTurk Participants once they complete their HIT. The second prompt directs participants to fill out their reasoning for their decisions. }
\label{fig:exit}
\end{figure*}

The experimental interface presented to MTurk Workers after acceptance of our HIT are shown in Figures \ref{fig:experiment_interface} to \ref{fig:exit}.

\subsection{Choosing $\lambda$ for Our Nash Approximation}
\begin{figure*}
\centering
\includegraphics[width=\textwidth]{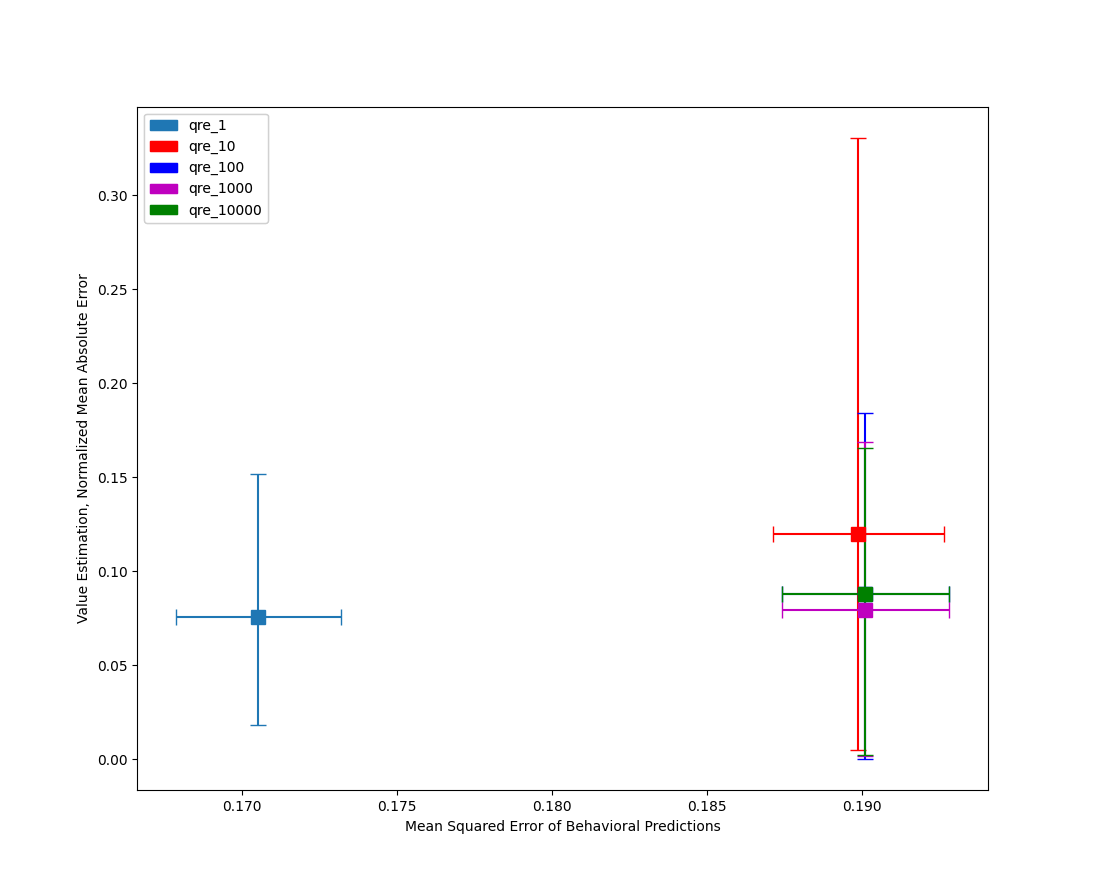}
\caption{QRE with fixed values of $\lambda$. As the error stabilizes around $\lambda = 100$, we use this as our \nashapprox}
\label{fig:nash_summary}
\end{figure*}

To select $\lambda$ for our Nash approximation, we compared the value estimates and likelihoods of behavioral predictions for several values of $\lambda$. We choose the lowest value of $\lambda$ at which both the value estimates and the likelihoods no longer change by increasing $\lambda$ further. Figure \ref{fig:nash_summary} shows this convergence in both value estimate and behavioral prediction.

\newpage
\subsection{Allocation Game Mapping Algorithm}
\label{apx:random_allocation}
To convert our payoff games to arbitrary allocation games, we use algorithm \ref{alg:random_allocation}.

\begin{algorithm}
\caption{Random allocation game generation algorithm}\label{alg:random_allocation}
\begin{algorithmic}
    \State Given set of payoff games $G$
    \State Given value $v^*$

    \For{$g \in G$}
        \For{$u(s_i, s_{-i}) \in g$}
        \State Sample $x \sim U(0, max(u(G))  /  v^* $
        \State Compute value $p$ where $p = u(s_i, s_{-i}) - x \times v$
        \State \Return $(x, p)$

        \EndFor
    \EndFor

\end{algorithmic}
\end{algorithm}

\newpage

\newpage
\section{Additional Figures and Tables}

\subsection{Behavioral Parameter Estimates}\label{section:behavioral_estimates}
This section gives additional information on the estimated behavioral parameters. Table \ref{tab:qch_ql4_tau} gives the Poisson mean parameter $\tau$ we back out for different \truevalue across treatments, and \ref{tab:qreplus_beta } gives the proportion $\beta$ of non-strategic agents.
\begin{table*}[h]\centering
\caption{Estimated $\tau$ when using \qchlinear, with $\tau$ indicating the rate parameter for a Poisson distribution specifying the proportion of agents of level $k$.}\label{tab:qch_ql4_tau}
\scriptsize
\begin{tabular}{cccc}\toprule
\truevalue &Combined Filtered &Combined Nonfiltered \\\midrule
5 &0.32654 (0.22824 0.42485) &0.48453 (0.32990 0.63915) \\
10 &0.40766 (0.27434 0.54098) &0.47658 (0.32487 0.62830) \\
20 &0.35799 (0.27176 0.44421) &0.39598 (0.24480 0.54717) \\
40 &0.41049 (0.35996 0.46102) &0.28461 (0.15282 0.41640) \\
80 &0.35470 (0.17743 0.53198) &0.50106 (0.32416 0.67796) \\
\bottomrule
\end{tabular}
\end{table*}

\begin{table*}[h]\centering
\caption{Estimated $\beta$ when using \qrep{QL4}, with $\beta$ indicating the proportion of agents who are non-strategic}\label{tab:qreplus_beta }
\scriptsize

\begin{tabular}{cccc}\toprule
\truevalue &Combined Filtered &Combined Nonfiltered \\\midrule
5 &0.70950 (0.65445 0.76456) &0.57524 (0.45547 0.69502) \\
10 &0.67130 (0.59895 0.74366) &0.56662 (0.46635 0.66688) \\
20 &0.68092 (0.63536 0.72647) &0.45589 (0.40195 0.50983) \\
40 &0.78786 (0.70963 0.86610) &0.56688 (0.52949 0.60427) \\
80 &0.73638 (0.65591 0.81685) &0.41873 (0.39231 0.44515) \\
\bottomrule
\end{tabular}

\end{table*}

Table \ref{tab:nodollar_estimates } shows the effect of not including payments within the allocation games. Even when using our best model in PQCH-QL4, the estimated values are incorrect, failing to scale to the correct value,  especially at lower values of \truevalue. This issue does not seem to be as pronounced in QRE, but the relative error is worse than allocation games containing payments.

\begin{table*}[ht]\centering
\caption{Raw value estimates when allocation games contain no payments for PQCH-QL4.  }\label{tab:nodollar_estimates }
\scriptsize\resizebox{\textwidth}{!}{
\begin{tabular}{lrrrrrr}\toprule
$v^*$ &5 &10 &20 &40 &80 \\\midrule
PQCH-QL4  &53.75, (0.63, 106.87) & 77.89, (21.59, 134.19) &44.38, (15.13, 73.64) & 50.98, (29.21, 72.74) & 104.07, (73.05, 135.08) \\
QRE &6.86, (6.79, 6.93) &11.54, (11.53, 11.55) &2.05, (1.93, 2.17) &33.99, (33.88, 34.1) &89.92, (89.83, 90) \\
\bottomrule
\end{tabular}
}
\end{table*}

\begin{table*}[!htp]\centering
\caption{Comparison of estimated \truevalue and $\lambda$ and $\beta$ behavioral parameters when finding the parameter(s) which maximize the likelihood (QRE Likelihood) vs. computing QRE against the empirical distribution (QRE Empirical). T-distributed confidence interval in brackets}\label{tab:qre_fixed_comparison}
\scriptsize
\begin{tabular}{ccccccc}\toprule
        \multicolumn{3}{r}{\textbf{QRE Likelihood}} & \multicolumn{3}{r}{\textbf{QRE Empirical}} \\
        \cmidrule(lr){2-4}\cmidrule{5-7}
$V^*$ & $\lambda$ & $\beta$ & $v$  & $\lambda$ & $\beta$  & $v$ \\\midrule
5 &0.465 (0.127) &0.869 (0.045) &4.857 (0.327) &0.244 (0.209) &0.534 (0.116) &4.922 ( 0.227) \\
10 &0.27 ( 0.102) &0.811 (0.081) &9.801 (0.555) &0.087 (0.103) &0.447 ( 0.089) &10.161 ( 0.474) \\
20 &0.487 ( 0.148) &0.817 (0.079) &19.857 (1.12) &0.106 (0.089) &0.554 ( 0.1) &19.629 (0.735) \\
40 &0.389 (0.147) &0.803 (0.075) &39.17 (1.879) &0.017 (0.006) &0.504 ( 0.085) &39.011 (1.454) \\
80 &0.442 (0.131) &0.878 (0.06) &79.194 (3.053) &0.094 ( 0.161) &0.471 ( 0.078) &78.326 (2.649) \\
\bottomrule
\end{tabular}
\end{table*}

\end{document}